\documentclass[fleqn,usenatbib]{mnras}

\usepackage[utf8]{inputenc}
\usepackage{graphicx}
\usepackage{amsmath}
\usepackage{amssymb}
\usepackage{txfonts}
\usepackage[caption=false]{subfig}

\newcommand\lensed{\textsc{Lensed}}

\renewcommand\vec{\bmath}

\title[{\rm Lensed:} forward reconstruction from strong lensing]{%
\textsc{Lensed:} a code for the forward reconstruction of lenses and sources from strong lensing observations
}

\author[Tessore, Bellagamba, Metcalf]{
Nicolas Tessore,$^{1,2}$\thanks{E-mail: \href{mailto:nicolas.tessore@manchester.ac.uk}{\nolinkurl{nicolas.tessore@manchester.ac.uk}}}
Fabio Bellagamba$^1$
and R. Benton Metcalf$^1$
\\
$^1$Department of Physics and Astronomy, Università di Bologna, viale Berti Pichat 6/2, 40127 Bologna, Italy \\
$^2$Jodrell Bank Centre for Astrophysics, University of Manchester, Alan Turing Building, Manchester M13 9PL, UK
}

\pubyear{2016}

\begin{document}
\label{firstpage}
\pagerange{\pageref{firstpage}--\pageref{lastpage}}
\maketitle

\begin{abstract}
Robust modelling of strong lensing systems is fundamental to exploit the information they contain about the distribution of matter in galaxies and clusters. In this work, we present \lensed, a new code which performs forward parametric modelling of strong lenses. \lensed\ takes advantage of a massively parallel ray-tracing kernel to perform the necessary calculations on a modern graphics processing unit (GPU). This makes the precise rendering of the background lensed sources much faster, and allows the simultaneous optimisation of tens of parameters for the selected model. With a single run, the code is able to obtain the full posterior probability distribution for the lens light, the mass distribution and the background source at the same time. \lensed\ is first tested on mock images which reproduce realistic space-based observations of lensing systems. In this way, we show that it is able to recover unbiased estimates of the lens parameters, even when the sources do not follow exactly the assumed model. Then, we apply it to a subsample of the SLACS lenses, in order to demonstrate its use on real data. The results generally agree with the literature, and highlight the flexibility and robustness of the algorithm.
\end{abstract}




\section{Introduction}

\footnotetext[1]{\url{http://glenco.github.io/lensed/}}

Strong gravitational lensing is now allowing researchers to address questions about the distribution of matter, both dark and light, within galaxy clusters and individual galaxies, that cannot be addressed in any other way.
The details of the distribution of mass in these systems have implications for such fundamental topics as the interactions of dark matter with itself and baryons, the mass of the dark matter particle and constraints on possible modifications of the theory of gravity, along with astrophysical questions such as how baryons segregate from dark matter and how the efficiency of star formation depends on the dark matter distribution.
Strong lenses are also allowing astronomers to detect some of the most distant objects ever seen and study galaxy formation at the cosmic dawn.
Future surveys such as \textit{Euclid} \citep{2011arXiv1110.3193L} are expected to increase the number of known lenses by one or two orders of magnitude, and the scientific potential is indeed great if these lenses can be put to good use.

At the heart of all strong lensing studies is the modelling of lenses based on astronomical images.
For observations with resolved sources, \citet*{2013NewAR..57....1L} gave an overview of currently available software.
Successful methods for lens reconstruction have been presented recently by \citet{2003ApJ...590..673W}, \citet{2005ApJ...623...31D}, \citet{2006MNRAS.371..983S}, \citet{2006ApJ...649..616P}, \citet{2009MNRAS.392..945V}, \citet{2011ApJ...734..104N}, \citet{2013ApJ...777....1B}, and \citet{2015MNRAS.452.2940N}.


Among the range of current lens reconstruction algorithms, \emph{forward reconstruction} methods calculate an expected image from a model of the lens and source.
By changing the parameters of the model, the forward reconstruction tries to minimise differences between the reconstruction and observation, which are usually evaluated by a chi-square test or similar.
These methods have been very successful in the reconstructions of recent galaxy-galaxy lensing surveys such as SLACS \citep{2006ApJ...638..703B,2008ApJ...682..964B,2009ApJ...705.1099A}.
Because both the deflection field and the surface brightness distribution are modelled, these methods use all available information in the observation.
A further advantage is the ability to use parametric models for background sources, thereby allowing a systematic study of the parameters of lensed galaxies that would otherwise be inaccessible \citep{2007ApJ...671.1196M,2011ApJ...734..104N,2013ApJ...777....1B}.
The difficulty of the forward method lies in the thorough exploration of the model space, while simultaneously allowing for enough freedom in the model to represent reality.


In this article, we present \lensed,\footnotemark\ a new implementation of the forward reconstruction method with focus on speed, quality of the lensing simulation, and statistical robustness of the results.
The code was developed with the following features and goals in mind:%
\begin{itemize}%
\item User friendliness -- We want the code to be widely and easily used by the scientific community. This includes easy importation of images, PSFs and masks.
\item Full posterior reconstruction -- We want the code to report statistically well-justified errors, including degeneracies, for the lens model parameters.
\item Insensitivity to initial parameters -- We want to make it unnecessary to guess the values of any parameters before beginning the fit.
\item Flexible modification of models -- We want users to be able to easily create any model they choose and use the code to constrain its parameters.
\item Simultaneous fitting of sources and lenses -- The number of sources and lenses can be chosen by the user. Sources include visible foreground components and the sky brightness, both of which are usually subtracted in a separate step before lens fitting.
\item GPU acceleration -- To achieve the necessary exploration of the parameter space in a reasonable period of time, the code must be fast.
\item Portability -- To allow for wide use, the code must be able to run on different platforms with minimal reconfiguration.
\end{itemize}
How we addressed each of these goals will be detailed below.
 
The outline of this article is as follows.
We describe the theory of the forward reconstruction method in section~\ref{sec:fwdrec}.
The particulars of our implementation of the method are given in section~\ref{sec:implementation}.
Section~\ref{sec:testing} contains the tests we perform on mock observations to verify and validate the performance of our code.
Finally, in section~\ref{sec:slacs} we use \lensed\ to analyse real space-based observations and compare the results to existing reconstructions.

\section{The Bayesian forward method}
\label{sec:fwdrec}

In this section, we wish to briefly develop the two fundamental parts of the forward method for lens reconstruction: simulation of the physical system -- lenses, sources, foregrounds etc. -- using an assumed model, and comparison of the predicted image with the true observation.
The latter task requires careful attention, because for any comparison, one must first define a measure of similarity.
Instead of using an \textit{ad hoc} criterion, we seek an objective metric for our reconstructions, which is derived in section~\ref{sec:comparison} in the form of a likelihood function for the model and data.
With this, we can perform a reconstruction using the forward method in a fully probabilistic, Bayesian approach.

\subsection{Simulating the model}
\label{sec:simulation}

The first part in modelling the physical system is to divide it into distinct planes on which there are two-dimensional lenses and sources.
Lenses act as deflectors of the line of sight, and sources contribute to the plane's surface brightness distribution.
This is the usual approach to lensing, and an introduction can be found e.g.\@ in the book by \citet*{2006glsw.conf.....M}.

We use a standard gravitational lens system consisting of one source plane and one image plane.
The lens acts on the image plane; it maps an apparent angular position $\vec x$ to the angular position $\vec y$ on the source plane according to the \emph{lens equation}
\begin{equation}\label{eq:lenseq}
    \vec y(\vec x) = \vec x - \vec\alpha(\vec x) \;. \end{equation}
The function $\vec\alpha$ is the \emph{deflection angle}, which for the purpose of this method characterises the lens completely.
Because of the deflection of light due to the lens, the surface brightness distribution~$I_S$ in the source plane is not directly observable.
Instead, the surface brightness distribution $I$ in the image plane is
\begin{equation}\label{eq:f}
    I(\vec x) = I_L(\vec x) + I_S(\vec y(\vec x)) \;,
\end{equation}
where $\vec y(\vec x)$ must be evaluated according to the lens equation~\eqref{eq:lenseq}, and $I_L$ is a contribution of (unlensed) light from the lens plane itself.

Using the total surface brightness $I(\vec x)$ of the image plane, it is possible to calculate the expected luminosity $L_i$, $i = 1, \dots, m$, for each of the $m$ pixels of the image, as the integral
\begin{equation}\label{eq:pixlum}
    L_i = \int_{P_i} \! I(\vec x) \, d\vec x
\end{equation}
of the surface brightness $I(\vec x)$ over the pixel area $P_i$.

In any real observation, the obtained image is degraded by the instrument and the measurement process.
This is usually modelled using a point-spread function (PSF), which is the observed image of an idealised point source.
The PSF is usually provided for a given instrument and observation.
In order to apply the PSF $P$, it is convolved with the physical surface brightness distribution $I$,
\begin{equation}
	I^*(\vec x)
    = \int \! P(\vec x - \vec x') \, I(\vec x') \, d\vec x'
    = \int \! P(-\vec r) \, I(\vec x + \vec r) \, d\vec r \;,
\end{equation}
to give the effective surface brightness distribution $I^*$ that is seen by the instrument.
The expectation values for the pixel luminosity in this case become
\begin{equation}\label{eq:psflum}
    L^*_{ij} = \int_{P_{ij}} \! I^*(\vec x) \, d\vec x \;,
\end{equation}
which is of the same form as expression~\eqref{eq:pixlum} without PSF.
In the next section, we will not treat the reconstruction with and without PSF separately; it is understood that the convolution must be performed and the latter expression used whenever a PSF is to be applied.

\subsection{Comparing model and observation}
\label{sec:comparison}

Having found the luminosity $L_i$ from \eqref{eq:pixlum} or \eqref{eq:psflum} for each pixel, the expected values of the model can be compared to the array of observed pixel values $d = \{ d_1, \dots, d_m \}$.
Because of the nature of the observational process, each individual $d_i$ is a random variate.
If the assumed model describes the physical process correctly, the mean of pixel value $d_i$ is given by the expected luminosity $L_i$ in suitably chosen units, i.e.
\begin{equation}
	\left\langle d_i \right\rangle = L_i \;.
\end{equation}
In addition to the values $d_i$, we require the variances
\begin{equation}
	s_i^2 = \operatorname{Var}(d_i)
\end{equation}
for each pixel.
These are sometimes provided with the observation, but often they are not and we must estimate the variance from the data itself as detailed in section~\ref{sec:variance}.
Then we can approximate the distribution of the pixel values by the normal distribution
\begin{equation}\label{eq:obsdist}
	d_i^{} \sim \mathcal{N}(L_i^{}, s_i^2) \;,
\end{equation}
given that our model is indeed the correct one.
Here we have made the implicit assumption that the ensemble of pixels is independently observed, meaning that the number of counts registered in one pixel does not depend on the counts for any other pixel.
This assumption does not preclude the application of a PSF, which is an effect that happens before the counts for a given pixel are registered and is taken into account by using~\eqref{eq:psflum}.
However, correlations between pixels can be (and most certainly are) introduced by the instruments or data reduction techniques such as drizzling \citep{2002PASP..114..144F}.
Our assumption therefore requires that these correlations are small, and that the variance for each pixel is enough to describe the statistics of the observation adequately.

With the variance $s_i^2$ fixed, either externally or from the data, we are now able to give the probability distribution function for observing the pixel value $d_i$ given an expected luminosity $L_i$, which is the Gaussian
\begin{equation}\label{eq:obspdf}
	P(d_i \mid L_i) = \frac{1}{\sqrt{2\pi \, s_i^2}} \, \exp\left\{ -\frac{1}{2} \frac{(d_i - L_i)^2}{s_i^2} \right\} \;.
\end{equation}
For the whole set of observed values $d = \left\{ d_1, \dots, d_m \right\}$ and expected luminosities $L = \left\{ L_1, \dots, L_m \right\}$, we then find the likelihood
\begin{equation}\label{eq:imgpdf}
	P(d \mid L) = \prod_{i = 1}^{m} P(d_i \mid L_i)
\end{equation}
of observing data $d$ given $L$ by the model.
As discussed before, the step from \eqref{eq:obspdf} to \eqref{eq:imgpdf} requires the assumption that the pixels are independent.

Finally, we are now able to build the parametric model which we want to reconstruct using likelihood~\eqref{eq:obspdf}.
Such a model can be comprised of multiple individual lens and source objects, which as a whole describe the deflection and surface brightness in terms of the $n$ parameters~$\xi = \left\{ \xi_1, \dots, \xi_n \right\}$.
For given values of~$\xi$, we can use expression~\eqref{eq:pixlum} or \eqref{eq:psflum} to calculate the numerical values of the expected pixel luminosities $L_i$ in our image, which thus become implicitly functions of the parameters~$\xi$ of the model.
Therefore, we can identify the likelihood of the parameter values and \eqref{eq:imgpdf}, and write
\begin{equation}\label{eq:parlike}
	P(d \mid \xi) \equiv P(d \mid L) = \prod_{i = 1}^{m} P(d_i \mid L_i)\;.
\end{equation}
Using Bayes' theorem, we subsequently invert the conditionality in likelihood~\eqref{eq:parlike} and find
\begin{equation}\label{eq:posterior}
	P(\xi \mid d) = \frac{P(d \mid \xi) \, P(\xi)}{P(d)}
\end{equation}
for the posterior probability $P(\xi \mid d)$ of the parameters $\xi$ given the observation $d$ and prior probabilities $P(\xi)$.
The constant of normalisation $P(d)$ is the Bayesian evidence, which can be calculated as the marginal distribution
\begin{equation}\label{eq:evidence}
	P(d) = \int \! P(d \mid \xi) \, P(\xi) \, d\xi
\end{equation}
and used for model selection.
Because the evidence~$P(d)$ is a global constant, many algorithms that sample posterior~\eqref{eq:posterior} do not require its knowledge.
This is in particular true for the class of Markov chain Monte Carlo (MCMC) samplers.

\section{Implementation}
\label{sec:implementation}

We will now give an explanation of our implementation of the forward reconstruction method as detailed in section~\ref{sec:fwdrec}.
A high-level algorithmic version could be the following:
\begin{enumerate}
\item Start with an image of observed pixel values $d_i$ and their observational variance $s_i^2$.
If no variance is provided, generate it from the data.
Apply mask when one is provided.
\item Build a model of the lens and sources, parametrised by a number of parameters~$\xi$ with prior probabilities~$P(\xi)$.
\item Pick a set of parameters $\xi$ from the prior $P(\xi)$.
\item For the chosen parameters, calculate the expected luminosities $L_i$ using \eqref{eq:pixlum}.
\item Calculate likelihood $P(d \mid \xi)$ of the model using \eqref{eq:obspdf} and \eqref{eq:imgpdf}.
\item Calculate the posterior probability $P(\xi \mid d)$ using the likelihood $P(d \mid \xi)$ and prior $P(\xi)$.
\item Repeat steps (iii) -- (vi) until the parameter space is sufficiently sampled.
\end{enumerate}
Even though the algorithm in this form is conceptually simple, there has so far been no general purpose standard implementation of the forward reconstruction method for lensing.
We think that there are two key obstacles that have to be overcome, which will be detailed in the following.

The first problem is step (iv) of the schematic algorithm laid out above:
A precise calculation of the expected luminosities \eqref{eq:pixlum} of the model requires a full ray-tracing simulation of the field of view to find the lensed surface brightness distribution, which must be subsequently integrated for each pixel of the image.
With modern space-based observations of galaxy-galaxy lensing events, the number of pixels that have to be calculated in this way routinely ranges from 10.000 to 100.000, making a numerical integration very costly.
A possible way around this problem is the ``poor man's ray tracing'' employed e.g. by \citet{2007ApJ...671.1196M}, where the simulated image samples the lensed surface brightness distribution at a fixed increased resolution.
Naturally, there are limitations to the amount of time one can save in the computations before the method breaks down, as the real and estimated pixel luminosities near the highly magnified parts of the image depend critically on the quality of the numerical integration.

The second problem of the forward method is step (vii), i.e.\@ the sampling of the parameter space.
For example, a reconstructed model might only fit some of the multiple images of a source at a time.
In order to fit the remaining images, an algorithm must be able to move away from ``good'' parameter values, because the posterior distribution is naturally multi-modal.
This rules out a standard maximum-likelihood approach for reconstruction, and requires a full sampling of the posterior instead.
There are also a number of inherent degeneracies: for example, to swap the major and minor axis of a nearly-circular ellipsoidal lens, the position angle must be changed by $\pm90\degr$, even though the models are physically similar.
Finally, the reconstruction is beset with strong correlations between parameters.
Some are inherent in the surface brightness distribution of sources, e.g.\@ between the position, ellipticity, and position angle.
Others arise from gravitational lensing, where observed features of an image can be attributed to both the light deflection by the lens and the intrinsic shape of the background source.
Many strategies exist to improve the sampling of the parameter space, e.g.\@ by taking known degeneracies into account.
Further speed up could be achieved by a decomposition of the parameters into ``fast'' and ``slow'' variables, where ``fast'' variables are updated more frequently while the slow variables are kept fixed.
This could help find e.g.\@ the magnitude of sources quickly, as pixel values can be linearly scaled for any given values of the remaining parameters.
Multi-modality can be tackled e.g.\@ by first scanning the full parameter space for global maximum-likelihood regions, which are subsequently analysed as usual \citep*{2015ApJ...813..102B}.

With \lensed, we provide an implementation of the forward method that tries to handle the above issues in full generality.
We believe that avoiding special treatment is the best way to achieve our stated goals of \emph{correctness of the simulation} and \emph{robustness of the results}.
The result is an algorithm built around a massively parallel ray-tracing kernel that performs the necessary calculations on a modern graphics processing unit (GPU).
Device code is written in \textsc{OpenCL} (a dialect of \textsc{C}), to ensure compatibility with commonly used devices and preserve the ability to compute on traditional CPU architectures when necessary or desirable (an overview of \textsc{OpenCL} was given by \citealp*{Stone:2010he}).
Our architecture has the added benefit that new source and lens models can be implemented quickly and without the need to recompile the code, opening up the way to purpose-built models for individual reconstructions.
The setup of the physical system, as well as the program itself, can be done entirely with one configuration file.

\subsection{Source integration}
\label{sec:integration}

The simulation of a given model consists of the calculation of the expected pixel luminosities~\eqref{eq:pixlum}, which must be done numerically.
Even though one cannot construct simple analytical models that capture the small-scale structure of real galaxies, it is nevertheless important that the integration is precise enough to not degrade the results of the reconstruction.
Crude methods can introduce extra correlations between the source and lens components:
For example, when using a single ray to determine the luminosity of a pixel, the lens might be coerced to optimally position the probe rays over the brightness distribution of the source.
This can lead to a bias in results for an individual reconstruction, especially when a small or no PSF is used.
We have found that the computational expense of performing reasonably precise integration of the pixel luminosities is often compensated by an increase in smoothness of the posterior distribution, which improves the overall sampling.

The pixel luminosities~\eqref{eq:pixlum} are thus integrated numerically by a weighted sum%
\begin{equation}\label{eq:intsum}
    L_i = \int_{P_i} \! I(\vec x) \, d\vec x \approx \sum_{k=1}^{N} w_k \, I(\vec x_k) \;,
\end{equation}
with weights $w_k$ and abscissae $\vec x_k \in P_i$ prescribed by some rule for numerical integration.
In case a PSF is to be applied, it is assumed to be provided as an array of $(2p+1) \times (2q+1)$ function values $\{ F_{-p,-q}, \dots, F_{0,0}, \dots, F_{p,q} \}$ centred on the $(0,0)$ element.
The PSF is normalised automatically.
After the pixel luminosities $L_{ij}$ (here indexed in two dimensions for simplicity) have been calculated, the integral~\eqref{eq:psflum} is replaced by the discrete convolution
\begin{equation}\label{eq:psfconvolve}
	L_{ij}^* = \sum_{k=-p}^{p}\sum_{l=-q}^{q} F_{-k,-l} \, L_{i+k,j+l} \;.
\end{equation}
This convolution is calculated directly on the device (GPU/CPU).
Special care has to be taken near the borders of the image, where we pad the array to the required size by repeating the closest pixel value $L_{ij}$.
As usual, no significant signal should be present in the border region of $p,q$ pixels when using a PSF.

For the integration rule, we selected the Cartesian square of a classical seven-point Gau\ss--Kronrod quadrature rule \citep{Patterson:1968fm} to be the default setting of \lensed.
The choice is based on the testing procedure and comparison of different rules shown in section~\ref{sec:testing}, and offers reasonable performance and good accuracy for the reconstruction of lenses as well as lensed background sources.
As usual for Gau\ss--Kronrod quadrature, an embedded three-point rule estimates the error of the integration for each pixel, which is saved with the output and can be used as a diagnostic for the quality of the numerical integration.

\subsection{Sampling of the parameter space}
\label{sec:sampling}

For a practical application, it is almost always infeasible to explore posterior distribution~\eqref{eq:posterior} with a classic MCMC method, due to the high dimensionality of typical models and the usually strong correlations between individual model parameters.
A more recent alternative to traditional MCMC samplers is the Nested Sampling algorithm given by \citet{2004AIPC..735..395S,Skilling:2006gi}.
We use the \textsc{MultiNest} library \citep{2008MNRAS.384..449F,2009MNRAS.398.1601F,2013arXiv1306.2144F}, which is an implementation and extension of the Nested Sampling algorithm that is well suited for working with the $10 < n < 50$ parameters, multiple modes, and correlations that typically arise in strong lensing reconstructions.

For a given problem, \textsc{MultiNest} requires the logarithm of the likelihood, which in our case~\eqref{eq:parlike} is
\begin{equation}\label{eq:logP}
	\log P(d \mid \xi) =
	 \sum_{i=1}^{m} -\frac{1}{2} \left\{\frac{(d_i - L_i)^2}{s_i^2} + \log(2\pi s_i^2) \right\} \;.
\end{equation}
Since the observational variance $s_i^2$ is fixed, we can use
\begin{equation}\label{eq:loglike}
	\operatorname{loglike}(d \mid \xi) = -\frac{1}{2} \sum_{i=1}^{m} w_i \left(d_i - L_i\right)^2 \;,
\end{equation}
to perform the required calculations, where we have introduced the weights $w_i = 1/s_i^2$.
We have dropped the second term from \eqref{eq:logP}, which amounts to an overall normalisation that has no bearing on the results.
An advantage of working with the weights $w_i$ instead of the variance $s_i^2$ is that a pixel can easily be masked by setting its weight to zero.
We further note that the sum in \eqref{eq:loglike} amounts to the usual $\chi^2$ term, which we can use to summarise the ability of the given parameter values $\xi$ to reconstruct the observed data.

The \textsc{MultiNest} algorithm offers many settings that can be used to find the right balance between speed and thoroughness of the sampling for a given task, and \lensed\ exposes all of them to the user.
We have tried to find reasonable default settings that work well in most typical situations, with a slightly larger emphasis on speed.
These defaults should however in no way be considered to be ideal.
We refer users to the relevant \textsc{MultiNest} documentation and invite them to find the combinations that work best for the specific problem at hand, as this often leads to significant improvements in speed, or quality, or -- occasionally -- both.

We point out that in addition to the posterior distribution of the parameters, \textsc{MultiNest} calculates the evidence~\eqref{eq:evidence}, which is a difficult problem in numerical Bayesian statistics.
For any given model of lenses and sources, the evidence is a global, parameter-independent value that quantifies, as the name suggests, the confidence that the model is indeed responsible for the observed lensing event.
It can thus be used to objectively compare different models that reconstruct the same observation.
A good introduction into the topic was given by \citet{Kass:1995eh}.

\subsection{Variance estimation}
\label{sec:variance}

In the common case where the observation does not provide an independent estimate of the variance $s_i^2$ for each pixel, we must estimate it from the data $d_i$.
When the lensing signal is sufficiently strong, we can estimate $s_i^2$ from the Poisson statistics of the photon flux.
First, we compute the total photon counts $k_i$ for each pixel as
\begin{equation}\label{eq:k_i}
	k_i = g_i \left( d_i + b_i \right) \;,
\end{equation}
where $g_i$ is an effective gain and $b_i$ is an eventual offset that was subtracted from the input data.
The effective gain is the conversion factor from the data units to photon counts; for example, given an observation in ADU per unit time, the effective gain is
\begin{equation}
	g_i = (\text{electronic gain}) \times (\text{exposure time}) \;.
\end{equation}
The offset $b_i$ is any constant that has been subtracted from the pixel values; this is often the estimated sky background.
Assuming that the photons arrive at the detector with some fixed rate given by the physical system being observed, the counts $k_i$ follow a Poisson distribution, and we can estimate their variance as
\begin{equation}
	\operatorname{Var}(k_i) = k_i = g_i \left( d_i + b_i \right) \;.
\end{equation}
Rewriting \eqref{eq:k_i} for $d_i$ and taking the variance, we find
\begin{equation}\label{eq:varest}
	s_i^2 =
	\operatorname{Var}(d_i)
	= g_i^{-2} \, \operatorname{Var}(k_i)
	= g_i^{-1} \left(d_i + b_i\right)
\end{equation}
as an estimator of the variance in case it was not independently provided by the observation.

Where the signal from astronomical sources is low, other sources of noise, such as those due to the read-out process, become significant. They are not considered in the statistical variance described above and could be more difficult to estimate.
Determining the noise level directly from the data, e.g.\@ from the RMS values of empty patches of sky, depends highly on the data processing pipeline and can easily bias the reconstruction.
Therefore, we do not include this option in the code, but advocate for careful external preparation of the weight maps when necessary, so that the particulars of the observation can be correctly taken into account.
\lensed\ still offers the possibility to include a multiplicative extra weight factor, for a quick modelling of other contributions to the noise.

\section{Testing}
\label{sec:testing}

We have created a set of tests using virtual observations to verify and validate the \lensed\ code in a controlled environment.
Each test contains a suite of 100 mock images of a galaxy-galaxy lensing event for a given lens and source model.
While the parameters of the lens are fixed for all images, the source parameters are randomly selected for each image.
By subsequently analysing all 100 mock images, we can separate the influence of the randomised source from the ability of \lensed\ to successfully reconstruct the fixed lens.

In our tests, we consider a parametric lens and two different kinds of source model.
In section~\ref{sec:lens-exact}, mock images containing sources with parametric profiles are used.
Since both the lens and the source can be modelled exactly in \lensed, this provides an ``upper bound'' on our reconstruction abilities, and we wish to verify the functionality of the code by recovering lens parameters without any significant bias.
In the following section~\ref{sec:lens-inexact}, we replace the parametric sources with more realistic ones based on observed galaxies.
Because these sources cannot be modelled exactly, we use this test to validate the capability of \lensed\ to reconstruct the lens even if the sources are not described perfectly by their assumed model.
The same data sets are used in sections~\ref{sec:source-exact} and~\ref{sec:source-inexact} to investigate the ability of \lensed\ to recover the parameters of sources.
Finally, in section~\ref{sec:rules} we show how the choice of integration rule influences the results of these tests.

In each test image, the lens is a singular isothermal ellipsoid \citep*{1994A&A...284..285K} located at the centre of the image, with position $x_L = y_L = 0\arcsec$, Einstein radius $r_L = 1.3624$~arcsec, axis ratio $q_L = 0.75$, and position angle~$\theta_L = 45\degr$ measured counter-clockwise between the major axis and $x$-axis. 
Assuming typical lens and source redshifts of $z_L = 0.2$ and $z_S = 1.0$, respectively, the chosen Einstein radius translates into a velocity dispersion of $\sigma_v = 250\ \mathrm{km\ s^{-1}}$ for the isothermal ellipsoid.\footnote{%
However, \lensed\ works entirely in apparent units, and no information about cosmological distances is needed or implicitly employed.
}

A foreground galaxy is hosted in the lensing potential.
Based on the majority of current galaxy-galaxy lensing observations, we assume it to be an early-type object, which is well parameterised by the profile of \citet{1948AnAp...11..247D}.
The effective radius of the light profile is assumed to be $r_H = 2$~arcsec, or about 1.5 Einstein radii, which is a reasonable ratio for typical lensing galaxies \citep[e.g.][]{2013ApJ...777...97S}.
The other parameter values are randomly assigned for each image, using an axis ratio $q_H$ between 0.6 and 0.9, and major axis angle $\theta_H$ between 25\degr\ and 65\degr, in order to maintain a certain alignment between light and mass profile.
The foreground galaxy has a fixed magnitude of 16.5 mag and is co-centred with the mass distribution.

The model of the background galaxy varies with the test we perform.
In all cases, the background source is randomly positioned inside a circle circumscribing the caustic of the lens.
This is done in order to observe the strong lensing features necessary for lens reconstruction.

\begin{figure}%
\includegraphics[width=\columnwidth]{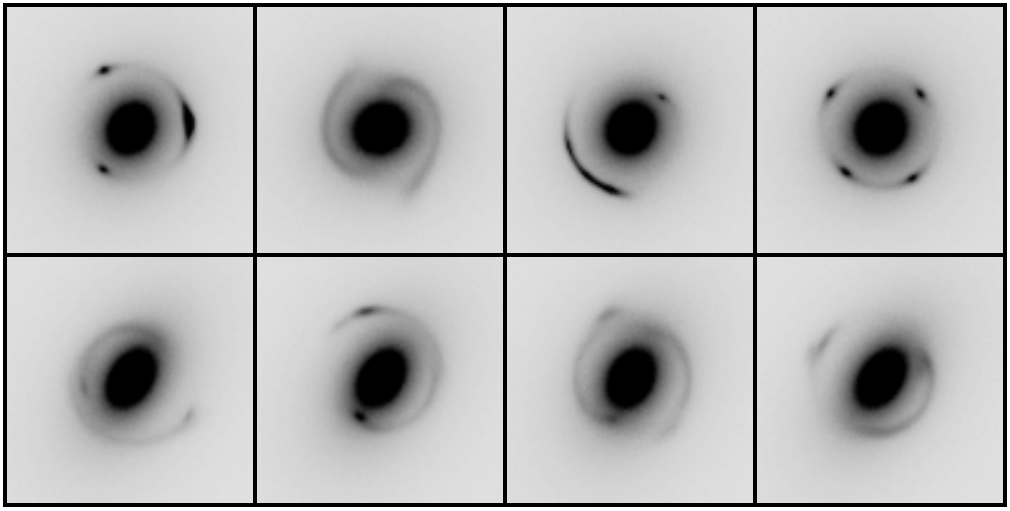}%
\caption{%
Sample images from the mock catalogs of 100 strong lenses using parametric Sérsic profiles (\emph{top}) and observed galaxies decomposed into a shapelet basis (\emph{bottom}) for the lensed galaxy.
The lens in each is a singular isothermal ellipsoid centred on the 6~arcsec image, with Einstein radius $r_L = 1.3624$~arcsec, axis ratio $q_L = 0.75$, and position angle $\theta_L = 45\degr$.
Background sources are positioned randomly within a disc circumscribing the caustic of the lens, with the remaining parameters randomised uniformly over their respective ranges.
}%
\label{fig:mock-images}%
\end{figure}

All mock observations in this section were realised with \textsc{Glamer}, a gravitational lensing simulator described by \citet*{2014MNRAS.445.1942M} and \citet*{2014MNRAS.445.1954P}.
Images are meant to realistically reproduce a space-based observation of the lensing systems.
We use a \textit{HST}-like configuration for the telescope properties, simulating a 2000~second exposure in the F814W band.
A typical PSF made with \textsc{TinyTim}\footnote{\url{http://tinytim.stsci.edu/}} is used for image synthesis and reconstruction.
These settings match the observations analysed in section~\ref{sec:slacs}.
The resulting images have a size of $120 \times 120$ pixels at a resolution of 0.05 arcsec/px, for a side length of 6~arcsec.
We choose to model this test on space-based observations because they are in general more difficult to reconstruct:
Ground-based observations of similarly-sized lenses contain fewer pixels, a larger PSF, and higher noise level, leading to a smoother likelihood with broader maxima.%

Fig.~\ref{fig:mock-images} shows a sample of the resulting virtual observations.
Each test set of 100 mock images contains the commonly found features of strong lensing such as rings, arcs, arclets, counter-images and crosses.
While this is not supposed to be a statistically representative sample of the lenses observed in any specific survey, it should nevertheless cover a wide range of the configurations for which this code was intended, observed under realistic conditions.

\subsection{Lens parameters from an exact model}
\label{sec:lens-exact}

For this test, the background galaxy follows an elliptical version of the Sérsic $r^{1/n}$ profile \citep{1963BAAA....6...41S,2005PASA...22..118G}, defined by
\begin{equation}\label{eq:sersic}
	I(R) = I_0 \exp\left\{-b_{n_S} \, (R/r_S)^{1/{n_S}}\right\} \;,
\end{equation}
where $R$ is the geometric mean radius
\begin{equation}\label{eq:radius}
	R = \sqrt{x^2 \, q + y^2 / q}
\end{equation}
of the ellipse passing through the point $x, y$.
The effective radius~$r_S$ therefore describes the geometric mean of the semi-major and semi-minor axes of the half-light ellipse.
The constant $I_0$ is chosen so that the source has a total magnitude of $\mathrm{mag}_S$.
This profile is then translated to position~$x_S, y_S$ and rotated by the position angle~$\theta_S$.
Since our aim is not to create a realistic model, but rather one that can be recovered perfectly, the mock images contain only a single background component.
Each image uses a different realisation of the uniformly random source parameters, with Sérsic index~$n_S$ between 0.5 and 8.0, effective radius~$r_S$ between 0.1 and 0.4~arcsec, axis ratio~$q_S$ between 0.1 and unity, and position angle~$\theta_S$ between 0\degr\ and 180\degr.
The range of values for the effective radius is compatible with observational results reported by \citet*{2015ApJS..219...15S}, while the ranges for the other parameters allow for a very general representation of possible source configurations.
The magnitude of the background galaxies before lensing is 23~mag in each case.

Each mock image is separately analysed with \lensed\ using the default settings.
In the spirit of this test, the model we assume is the same one used to generate the mock images: the lens is a singular isothermal ellipsoid, the foreground galaxy follows a de Vaucouleurs profile, and the background galaxy follows a Sérsic profile.
In total, the model contains $5 + 6 + 7 = 18$ parameters.
None of the parameters, including lens and source positions, are fixed or otherwise constrained.
We use uniform prior distributions that span the whole range of possible parameter values.

On a current workstation computer, the reconstruction of the full set of 100 images completes in a few hours.
In all cases but one, the recovered maximum-likelihood lens parameters are very close to the input values.
There was a sole outlier (Fig.~\ref{fig:mock-images}, top row, first column), which returns a minimal $\chi^2/\mathrm{dof} $ value of 6.3, indicating that the sampling of the parameter space terminated before the true maximum-likelihood region was found.
Further inspection of this system shows no particular oddities with respect to the total sample, and we therefore conclude that the premature termination is a numerical fluke.
Since the reconstruction involves a certain amount of randomness, such behaviour is occasionally expected, and can usually be detected by the high value of $\chi^2/\textrm{dof}$.
A subsequent analysis of this case with settings that result in a more thorough sampling of the parameter space (increased number of live points or reduced tolerance) aligns the results with the rest of the set.

\begin{figure}
\includegraphics[width=\linewidth, trim=10pt 10pt 10pt 10pt, clip]{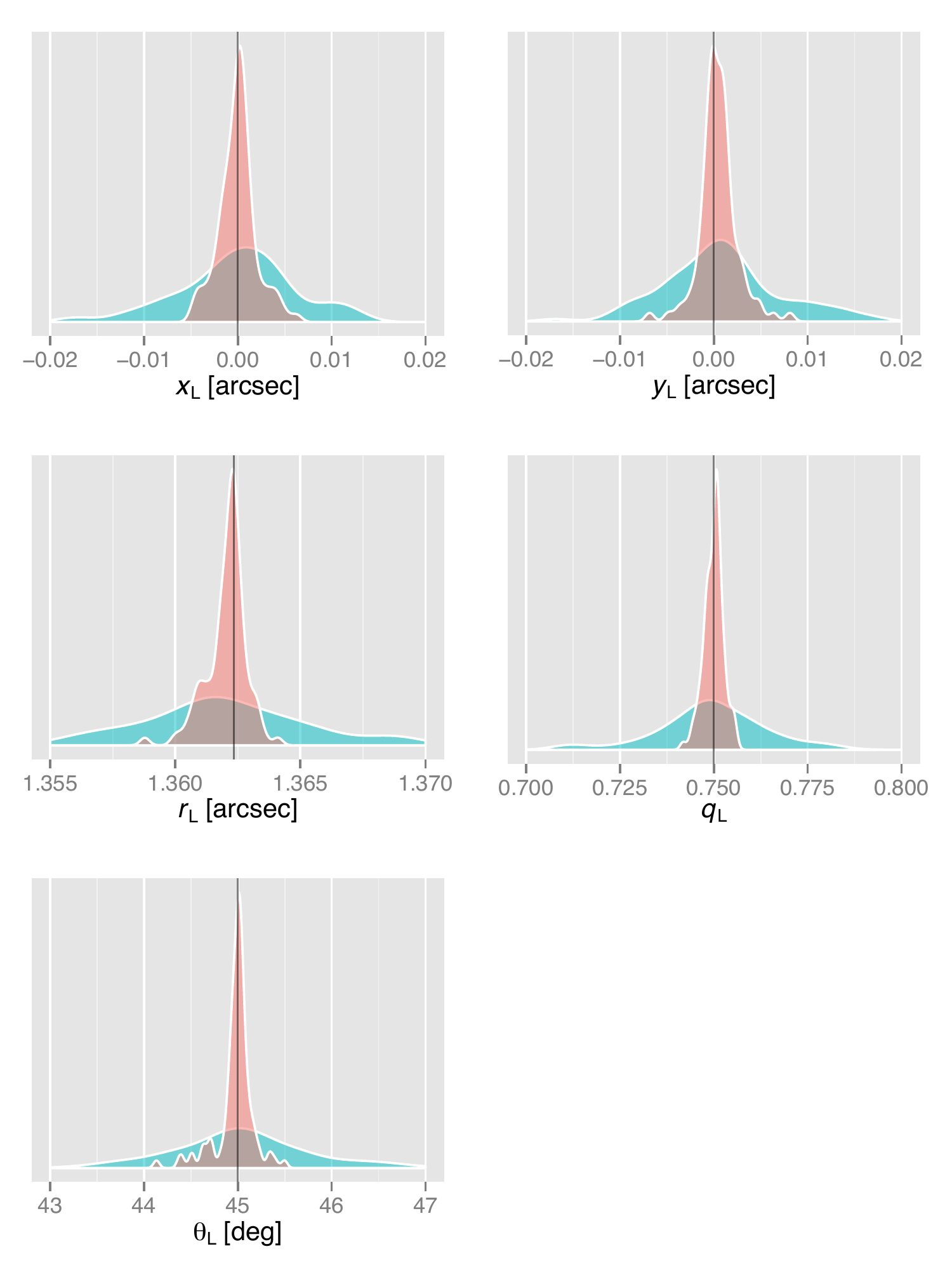}%
\caption{%
Posterior distributions of the lens parameters using an exact (\emph{red}) and inexact (\emph{blue}) model for the background sources, see text.
Shown are the position~$x_L$,~$y_L$, Einstein radius~$r_L$, axis ratio~$q_L$, and position angle~$\theta_L$ of a singular isothermal ellipsoid.
Input parameter values are indicated by a vertical line (\emph{black}).
Results are marginalised over 100 \textit{HST}-like mock observations with randomised sources.
}%
\label{fig:lens-parameters}%
\end{figure}

\begin{table}
\centering%
\caption[Constraining power on lens parameters]{%
Results of the reconstruction of mock images using an exact and inexact model for the background sources.
Shown are the parameters for position~$x_L$,~$y_L$, Einstein radius~$r_L$, axis ratio~$q_L$, and position angle~$\theta_L$.
The quoted values are the sample mean and standard deviation of the marginal distribution over 100 randomised realisations of the lens system, as shown in Fig.~\ref{fig:lens-parameters}
}%
\label{tab:lens-parameters}%
\setlength\tabcolsep{0.8\tabcolsep}%
\begin{tabular}{lccc}%
\hline
parameter & input value & exact source model & inexact source model \\
\hline
$x_L$ [arcsec]     & $0$       & $-0.0000 \pm 0.0020$  & $-0.001 \pm 0.011$  \\
$y_L$ [arcsec]     & $0$       &  $0.0005 \pm 0.0021$  & $-0.001 \pm 0.013$  \\
$r_L$ [arcsec]     & $1.36235$ & $1.36205 \pm 0.00080$ & $1.3619 \pm 0.0059$ \\
$q_L$              & $0.75$    &  $0.7499 \pm 0.0026$  &  $0.750 \pm 0.016$  \\
$\theta_L$ [\degr] & $45$      &   $44.97 \pm 0.20$    &   $45.2 \pm 1.7$    \\
\hline
\end{tabular}%
\end{table}

The distribution of the reconstructed lens parameters over the set of mock observations are shown in Fig.~\ref{fig:lens-parameters}.
No systematic bias in the outcome is apparent, and the results are tightly distributed around the true values.
This is substantiated in Table~\ref{tab:lens-parameters}, which contains the sample mean and standard deviation of the recovered lens parameters over the set of all 100 mock images.
The accuracy of the results, i.e. the vicinity of true value and average reconstructed value, is remarkably high.
The precision of the results, i.e. the scatter of the recovered values about the mean, is good as well:
position, Einstein radius, axis ratio, and orientation are constrained far below the pixel, per cent, and degree level, respectively.

This basic test confirms that \lensed\ is able to recover the lens in a controlled setting without introducing systematic biases.
Using the true model for reconstruction, we are able to recover the lens parameters almost identically.
In the next test, it will be shown how the constraints on lens parameters get widened once the background source model is no longer perfectly matched.

\subsection{Lens parameters from an inexact model}
\label{sec:lens-inexact}

A second set of images is made, in which background sources do not follow an analytical profile.
Instead, they are extracted from a catalogue of observed galaxies in the Hubble Ultra Deep Field \citep{2006AJ....132..926C} and subsequently decomposed into the shapelet basis functions \citep{2003MNRAS.338...35R}, using a code described by \citet*{2007A&A...463.1215M}.
Shapelet functions form a complete and orthonormal basis which is suited for the extraction and the manipulation of galaxy images, and are described in detail by e.g. \citet{2008A&A...482..403M}.

Only galaxies with a redshift between 0.8 and 1.5 are selected from the Ultra Deep Field catalogue, to ensure a sample of galaxy shapes consistent with an assumed source redshift of $z_S = 1.0$.
Moreover, by making a cut in apparent magnitude at 27 mag, we filter out galaxies that were observed with low S/N.
Each object is appropriately rescaled to have a (pre-lensing) magnitude of 23 mag and the same apparent size it would have had at redshift $z_S = 1.0$.

In this test, the lensed galaxies have a variety of shapes and substructures that cannot be modelled exactly.
Our goal is to check in how far this mismatch influences our ability to recover the lens parameters.
For the reconstruction, we use the same model as in section~\ref{sec:lens-exact}.
As before, all priors are chosen to be uniform over the whole range of possible parameter values.

The resulting parameter distributions of this test are shown in Fig.~\ref{fig:lens-parameters}.
The loss in precision due to the inexact nature of the reconstruction is clearly visible in the increased scatter of the recovered parameters.
Still, the distribution of the results, while not exactly Gaussian, is remarkably symmetric and centred on the true values.
As shown in Table~\ref{tab:lens-parameters}, the accuracy for the whole sample of 100 reconstructions remains far below the respective scales of one pixel, per cent, or degree.
Even though the standard deviation of the individual parameters has increased five- to tenfold over the exact case, the precision is still excellent, with constraints on the position and Einstein radius at the intrapixel level, the axis ratio at the per cent scale, and the orientation correct to within two degrees.
We have checked that the precision of the results can be increased using higher quality settings for the algorithm, although not to the same extent as in the exact case.

With this second test, we have shown that \lensed\ retains the ability to recover our given lens parameters even when the background galaxies cannot be modelled exactly.
Shapelet sources are diffuse, usually with multiple peaks and troughs in the brightness distribution, and are generally only poorly fit by a single Sérsic component.
Nonetheless, we were able to use \lensed\ to constrain the parameters of the given SIE lens model to a very high degree.
As we expect the reconstruction of real observations to be limited first and foremost by the model uncertainty due to an insufficient description of the mass distribution, we conclude that the effects of overly simplified sources are only secondary.
We are thus confident that the reconstructions of real observations, such as those presented in section~\ref{sec:slacs}, are not limited by our implementation of the forward method.

\subsection{Source parameters from an exact model}

\label{sec:source-exact}

\begin{figure}
\includegraphics[width=\linewidth, trim=10pt 10pt 10pt 10pt, clip]{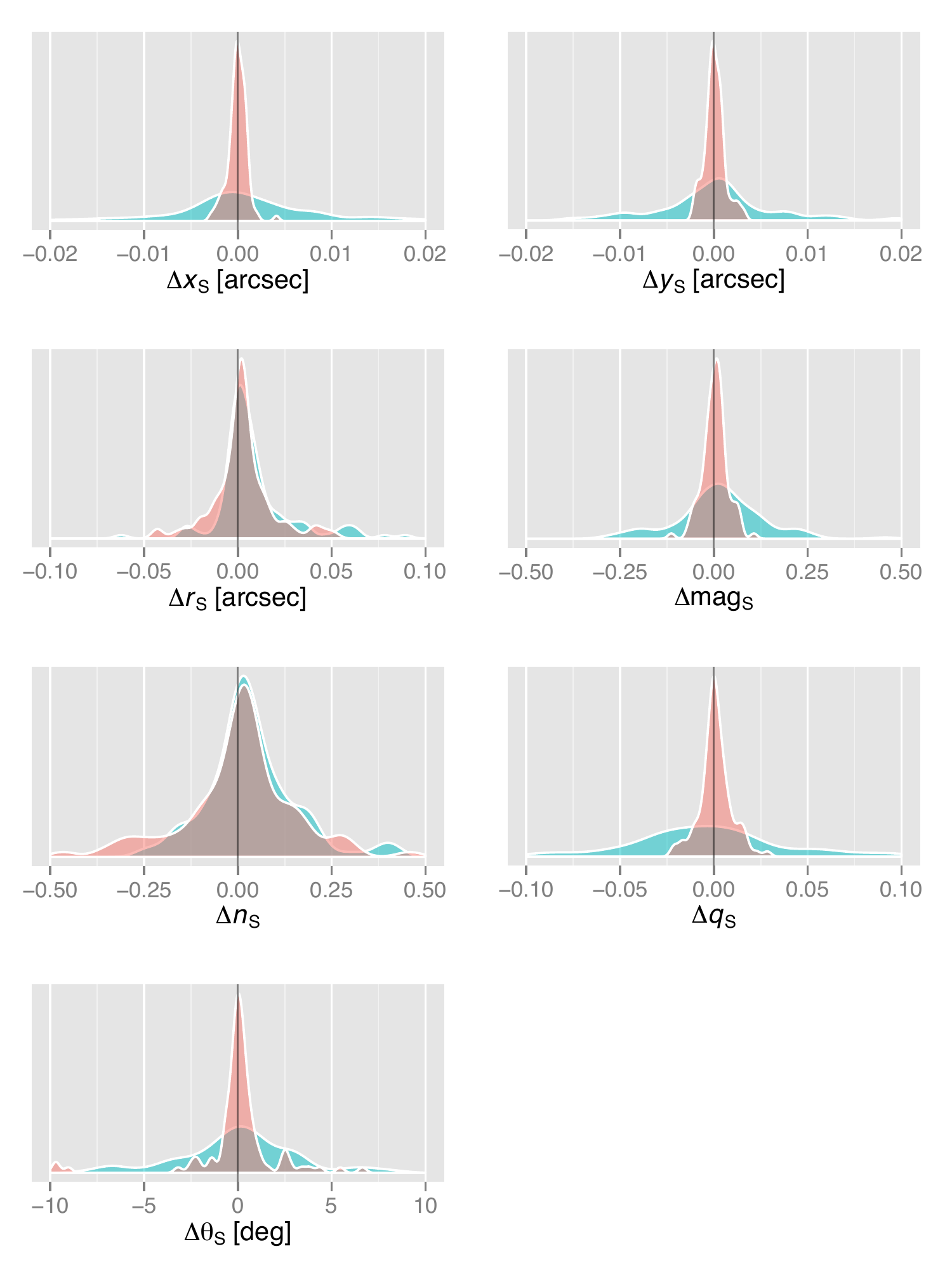}%
\caption{%
Posterior distributions of the source parameters using an exact (\emph{red}) and inexact (\emph{blue}) model for the background sources.
Shown are the errors of the reconstruction $\Delta = \text{(recovered)} - \text{(input)}$ for the recovered position~$x_S$,~$y_S$, effective radius~$r_S$, magnitude~$\mathrm{mag}_S$, S\'ersic index~$n_S$, axis ratio~$q_S$, and position angle~$\theta_S$ of the lensed sources, which are modelled in both cases with a S\'ersic profile.
Input parameters for the inexact sources were found as explained in the text. 
Results were obtained using the \textsc{Lensed} algorithm on \textit{HST}-like mock observations.
The results are marginalised over ensembles of observations with randomised sources, see text.
}%
\label{fig:source-parameters}%
\end{figure}

\begin{table}
\centering%
\caption{%
Constraining power of \lensed\ for the parameters of background sources.
Shown are results for the errors $\Delta$ in the recovered values, which are expected to vanish.
The corresponding distributions are shown in Fig.~\ref{fig:source-parameters}.
}%
\label{tab:source-parameters}%
\begin{tabular}{lcc}%
\hline
parameter & exact source model & inexact source model \\
\hline
$\Delta x_S$ [arcsec]    & $-0.0000 \pm 0.0010$ & $-0.002 \pm 0.027$ \\
$\Delta y_S$ [arcsec]    &  $0.0001 \pm 0.0011$ & $-0.002 \pm 0.029$ \\
$\Delta r_S$ [arcsec]    &   $0.003 \pm 0.019$  &  $0.008 \pm 0.034$ \\
$\Delta\mathrm{mag}_S$   &   $0.003 \pm 0.033$  &   $0.03 \pm 0.19$  \\
$\Delta n_S$             &    $0.01 \pm 0.25$   &   $0.07 \pm 0.25$  \\
$\Delta q_S$             &  $0.0009 \pm 0.0085$ & $-0.005 \pm 0.046$ \\
$\Delta\theta_L$ [\degr] &      $-1 \pm 14$     &      $1 \pm 15$    \\
\hline
\end{tabular}%
\end{table}

In many studies the gravitational lens is used as a serendipitous
telescope through which a  faint, distant high-redshift source can be observed \citep{2007ApJ...671.1196M,2011ApJ...734..104N,2013ApJ...777....1B}.
For these cases, the reconstructions of the source becomes as
important as the reconstruction of the lens, and both must be
performed simultaneously.  While it was shown in the previous tests
that the lens reconstruction is robust even when the sources are not
perfectly represented by the source model, the same is not necessarily true for the reconstruction of source parameters under the influence of lensing, and the question merits further testing.

The first step is again validating that \lensed\ is able to
reconstruct source parameters given an exact source model.
For this, the same ensemble of mock observations are used.  Here
instead of using the parameters $\xi$ themselves, the error of the
reconstructed parameters, $\Delta\xi = \xi - \xi_0$, is considered, so that all realisations have a common expectation value $\operatorname{E}[\Delta\xi] = 0$.
The results are shown in Fig.~\ref{fig:source-parameters}, and a summary is given in Table~\ref{tab:source-parameters}.
Photometric precision in this case of an exact model is 0.07~mag at the $2\sigma$ level, and the effective radius~$r_S$ of the sources is constrained to 12\% precision at the $2\sigma$ level.
The least constrained parameter is the Sérsic index~$n_S$, with a  relative uncertainty of 10\% at the $2\sigma$ level due to the general difficulty of recovering highly peaked sources (i.e.\@ those with $n_S > 2$).
Finally, the position angle~$\theta_S$ is very well reconstructed when the source is clearly elliptical, up to axis ratios of $q_S \approx 0.85$, while for almost circular sources, the position angle is naturally unconstrained.

Overall, \lensed\ can reliably reconstruct parametric sources which have been distorted by an unknown lens that is simultaneously reconstructed.
This is perhaps not a great surprise, but it is still a good validation of the fact that degeneracies and correlations between lens and source model do not restrict our ability of recovering source parameters.
The next step is to test whether this remains true when the sources are no longer described perfectly by the model used for their reconstruction.

\subsection{Source parameters from an inexact model}

\label{sec:source-inexact}

When the true source is not perfectly described by the input model, we would still like to be confident that the reconstruction does not find systematically different  source parameters due to some interaction between the deflection and the source profile. However, in this case assessing the quality of the recovered source is more difficult, because it is not obvious how to compute the input S\'ersic parameters of the sources. 

As described in section~\ref{sec:lens-inexact}, the objects used in this test are real objects, which have been decomposed in shapelet basis functions. This means that they have an analytical profile which, in theory, extends out to infinity. It is not straightforward to define the best S\'ersic parameters for this kind of object. In order to perform the fit to a S\'ersic profile, it is necessary to produce a map of uncertainties that act as inverse weights in the fitting procedure. In practise, this is equivalent to producing mock observations of the objects, and analysing them with a reconstruction code such as the one described in this paper. We would like these S\'ersic "input parameters" to be similar to what one would get if the source were fit no lens, high signal-to-noise, no pixelisation and no PSF blurring.

To this end, we produced virtual observations analogous to the ones presented in the beginning of this section for the lensing events. The only difference stands in the application of a uniform magnification that has the same total magnification $|\mu|_\mathrm{tot}$ as in the lensed case. The resulting images have high signal-to-noise values, just as the lensed images we produced. Moreover, they are not pixelated or unresolved with respect to the point-spread function, because the original shapelet decompositions were derived from HST data, and thus they cannot contain any information on scales smaller than one of the original pixels. This scale, due the magnification applied, corresponds to $\sim 10$ pixels in the final images, significantly larger than the PSF smoothing kernel. Using the same total magnification as in the lensing case also gives us confidence that the same region of the galaxy will be typically dominant over the noise, and significant for the extraction of the S\'ersic parameters.

Once the equivalent S\'ersic parameters of our sample of non-parametric shapelet sources have been found, the recovered and expected parameters can be compared as in the case of parametric sources.
The resulting marginal distributions are shown in Fig.~\ref{fig:source-parameters}.
While the shape (position~$x_S$,~$y_S$, axis ratio~$q_S$, position
angle~$\theta_S$) and the luminosity (magnitude~$\mathrm{mag}_S$) of
the sources are less constrained in comparison to the fully parametric case, the constraints on the parameters of the S\'ersic profile (effective radius~$r_S$ and S\'ersic index~$n_S$) remain practically unchanged.
This is substantiated in Table~\ref{tab:source-parameters}, which shows only a minor increase in variance for these parameters.


\subsection{Comparison of integration rules}

\label{sec:rules}

\begin{figure}%
\centering%
\includegraphics[width=\linewidth, trim=10pt 10pt 10pt 10pt, clip]{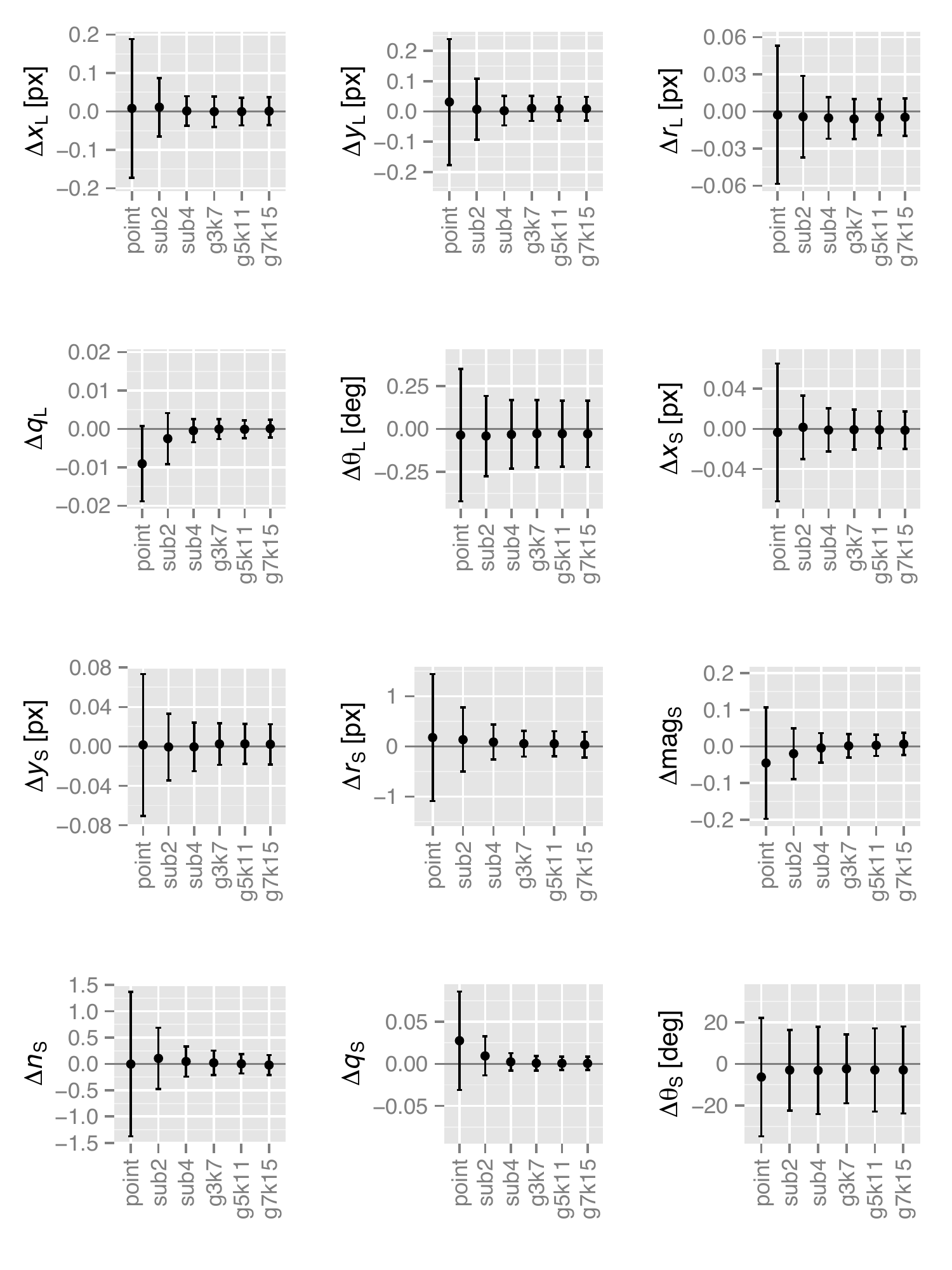}%
\caption{%
Comparison of lens reconstruction results for different integration rules.
Shown are the mean and standard deviations for the errors $\Delta$ over the sample of 100 reconstructions as a function of the chosen integration rule.
For a list of the tested integration rules, as well as the parameters, see text.
}%
\label{fig:integration}%
\end{figure}

To conclude the set of tests, it is briefly investigated how the results presented here are influenced by the choice of integration rule for the simulation of the model (cf.\@ section~\ref{sec:integration}).
While it is important that errors in the calculation of the expected image are small enough not to degrade the results of the lens reconstruction, there is no need to waste time on a perfect integration when it can be shown that this does not improve the quality of the results.
As usual in numerics, experimentation and testing is required to find the right choice for the task at hand, and an example is shown here.

The first test of this section can be repeated using different integration rules, in order to observe their influence on the recovered parameters.
The results of such an experiment using the exact test set is shown in Fig.~\ref{fig:integration}.
The evaluated integration rules are
\begin{itemize}
\item \texttt{point} -- a single point at the centre of each pixel,
\item \texttt{sub2} -- $2 \times 2$ subsampling of the pixels,
\item \texttt{sub4} -- $4 \times 4$ subsampling of the pixels,
\item \texttt{g3k7} -- Gau\ss--Kronrod quadrature rule with $7 \times 7$ points,
\item \texttt{g5k11} -- Gau\ss--Kronrod quadrature rule with $11 \times 11$ points,
\item \texttt{g7k15} -- Gau\ss--Kronrod quadrature rule with $15 \times 15$ points.
\end{itemize}
As before, the model parameters for the lens are position~$x_L, y_L$, scale radius~$r_L$, axis ratio~$q_L$ and position angle~$\mathrm{PA}_L$, while the source has position~$x_S, y_S$, effective radius~$r_S$, magnitude~$\textrm{mag}_S$, S\'ersic index~$n_S$, axis ratio~$q_S$, and position angle~$\theta_S$.
The results show a clear convergence of the results for the Gau\ss--Kronrod rules, while $4 \times 4$ subsampling has similar accuracy when only the lens parameters are considered.
Given the substantially lower number of integration points per pixel (16 instead of $\ge 49$), we conclude that $4 \times 4$ subsampling is the integration rule of choice for lens reconstruction, at least for \textit{HST}-like data using a realistic PSF at the same resolution.
When source parameters are required, one of the Gau\ss--Kronrod rules might be preferable in order to not bias the results, and we therefore select the $7 \times 7$-point rule as the default in our implementation.
However, rule selection is flexible in \lensed, and we encourage users to experiment.

\section{Application to real data}
\label{sec:slacs}

In order to show the feasibility of the analysis of real images with \lensed, we consider a selection of lenses from the Sloan Lens ACS Survey (SLACS) catalogue \citep{2006ApJ...638..703B,2008ApJ...682..964B}.
SLACS was a project that aimed to confirm with space-based observations lenses which were detected from ground-based SDSS spectroscopy.
The final catalogue \citep{2010ApJ...724..511A} comprises 85 confirmed lenses and 13 likely ones.
Analysing the whole catalogue is outside the scope of this work, and we limit ourselves to a demonstration of the capabilities of \lensed\ on real observations.
To this end, we analyse the first 8 observations of \textit{HST} proposal 10886 in the F814W band that are available on the Hubble Legacy Archive.
These SLACS lenses were originally investigated by
\citet{2008ApJ...682..964B}, who modelled 7 of the systems, with the
remaining system not modelled due to the presence of a companion to
the lens galaxy  (see Table~\ref{tab:slacs-results}).%

The base model for this analysis consists of a flat sky component (to account for imperfections in the process of sky subtraction), a Sérsic profile for the host galaxy, a SIE lens and a Sérsic profile for the background galaxy.
When the reconstruction with this model is not satisfactory (high $\chi^2$), we introduce additional Sérsic components for the foreground or background galaxies.
For this process, we consider a model detailed enough if there are no longer any systematic differences between the observation and reconstruction, leading to just one or two components per visible object.
In a more thorough analysis, one might choose the number of components using some objective criterion, and also impose a prior on their distribution \citep{2011MNRAS.412.2521B}.
Since \citet{2008ApJ...682..964B} forced the centre of the mass distribution to coincide with the centre of the surface brightness distribution for all lenses, we restrict ourselves to the same assumption for the sake of comparison.
We emphasise that this restriction is not necessary with our approach, as demonstrated in section \ref{sec:testing}.
All other parameters are left unconstrained.
In those cases where we need to exclude additional visible objects close to the lens from the analysis, we provide \lensed\ with a suitable mask of the uninteresting regions.

\begin{table*}%
\centering%
\caption{%
Estimated parameters for the SIE lens models of 8 SLACS lenses from \textit{HST} proposal 10886.
The table compares the mean and standard deviation obtained from \lensed\ with the SLACS lens modelling results published by \citet{2008ApJ...682..964B}.
The columns $N_\mathrm{fg}$ and $N_\mathrm{src}$ give the number of components used to model the foreground and background galaxies, respectively.
East-of-north angles from the original results have been changed to match our definition.
The dotted fields indicate that the lens was not originally analysed.
}%
\label{tab:slacs-results}%
\begin{tabular}{lccccccccc}
\hline
& \multicolumn{5}{c}{\hrulefill~~~\lensed~~~\hrulefill}
& \multicolumn{4}{c}{\hrulefill~~~SLACS~~~\hrulefill} \\
lens name & $b_\mathrm{SIE}$~[arcsec] & $q$ & P.A.~[\degr] & $N_\mathrm{fg}$ & $N_\mathrm{src}$ & $b_\mathrm{SIE}$~[arcsec] & $q$ & P.A.~[\degr] & $N_\mathrm{src}$ \\
\hline
SDSS J0029$-$0055    & $0.9603 \pm 0.0010$ & $0.8824 \pm 0.0020$ & $113.09 \pm  0.45$ & 1 & 1 & $0.96$ & $0.89$ & $115.4$ &     2 \\
SDSS J0252$+$0039    & $1.0244 \pm 0.0006$ & $0.9226 \pm 0.0011$ & $ 16.07 \pm  0.47$ & 1 & 1 & $1.04$ & $0.93$ &  $16.2$ &     3 \\
SDSS J0330$-$0020    & $1.1012 \pm 0.0008$ & $0.8119 \pm 0.0032$ & $ 24.09 \pm  0.24$ & 1 & 1 & $1.10$ & $0.81$ &  $23.2$ &     3 \\
SDSS J0728$+$3835    & $1.2344 \pm 0.0014$ & $0.8958 \pm 0.0027$ & $159.80 \pm  0.62$ & 2 & 2 & $1.25$ & $0.85$ & $157.6$ &     4 \\
SDSS J0808$+$4706(a) & $1.1699 \pm 0.0022$ & $0.7991 \pm 0.0019$ & $  9.37 \pm  0.41$ & 2 & 1 &  \dots &  \dots &   \dots & \dots \\
SDSS J0808$+$4706(b) & $0.3335 \pm 0.0045$ & $0.9943 \pm 0.0024$ & $111.36 \pm 10.82$ & 1 & 1 &  \dots &  \dots &   \dots & \dots \\
SDSS J0822$+$2652    & $1.1081 \pm 0.0037$ & $0.9577 \pm 0.0095$ & $ 99.14 \pm  7.39$ & 1 & 2 & $1.17$ & $0.88$ & $158.2$ &     2 \\
SDSS J0841$+$3824    & $1.3850 \pm 0.0024$ & $0.7831 \pm 0.0019$ & $171.42 \pm  1.02$ & 2 & 2 & $1.41$ & $0.79$ &   $1.4$ &     2 \\
SDSS J0903$+$4116    & $1.2911 \pm 0.0022$ & $0.8784 \pm 0.0038$ & $ 68.04 \pm  1.08$ & 1 & 1 & $1.29$ & $0.90$ &  $71.3$ &     2 \\
\hline
\end{tabular}%
\end{table*}

\begin{figure*}%
\subfloat[SDSS J0029$-$0055]{%
\includegraphics[width=\columnwidth]{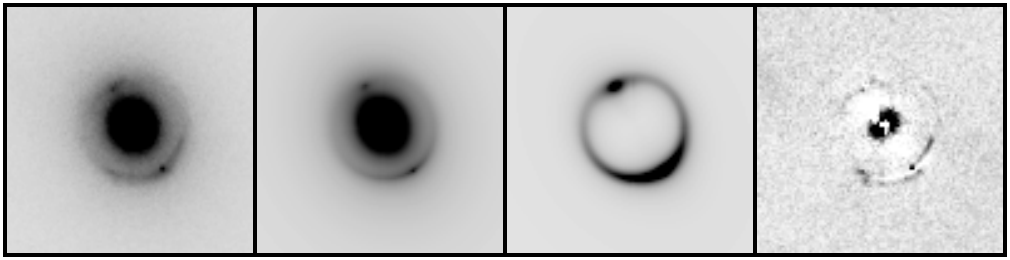}%
}%
\hfill%
\subfloat[SDSS J0252$+$0039]{%
\includegraphics[width=\columnwidth]{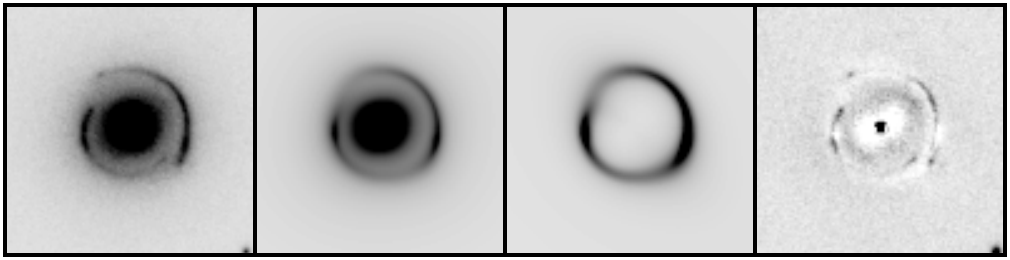}%
}%
\\%
\subfloat[SDSS J0330$-$0020]{%
\includegraphics[width=\columnwidth]{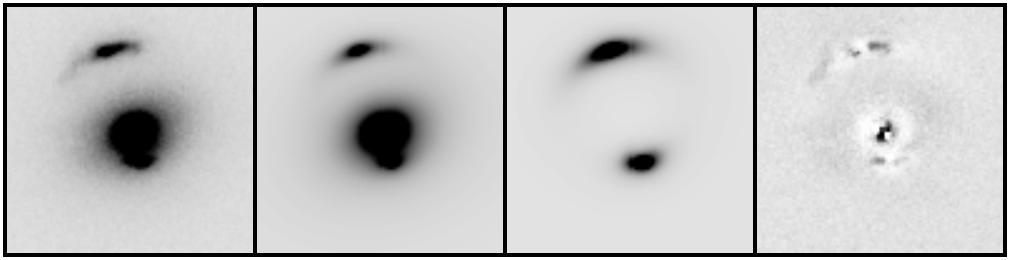}%
}%
\hfill%
\subfloat[SDSS J0728$+$3835]{%
\includegraphics[width=\columnwidth]{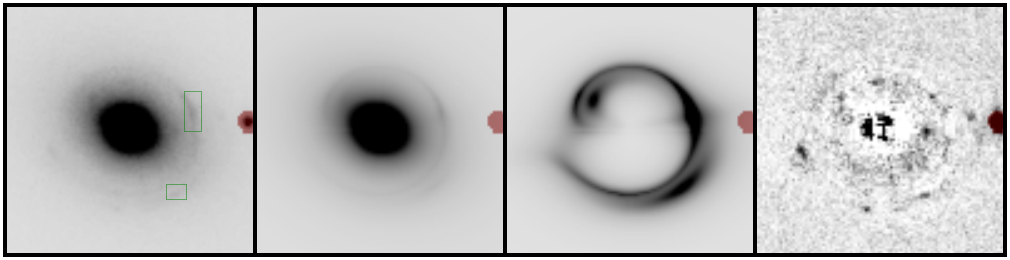}%
}%
\\%
\subfloat[SDSS J0808$+$4706\label{fig:J0808+4706}]{%
\includegraphics[width=\columnwidth]{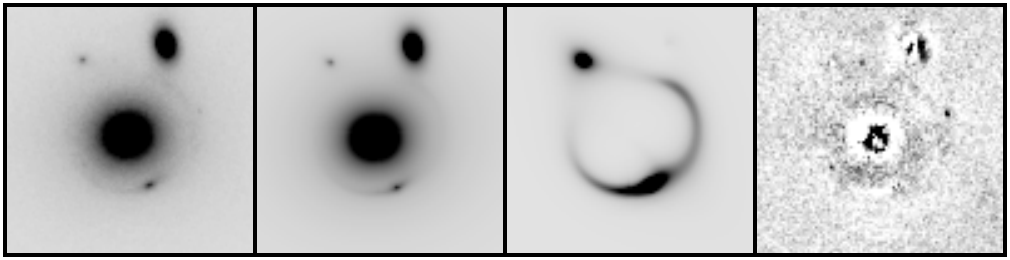}%
}%
\hfill%
\subfloat[SDSS J0822$+$2652]{%
\includegraphics[width=\columnwidth]{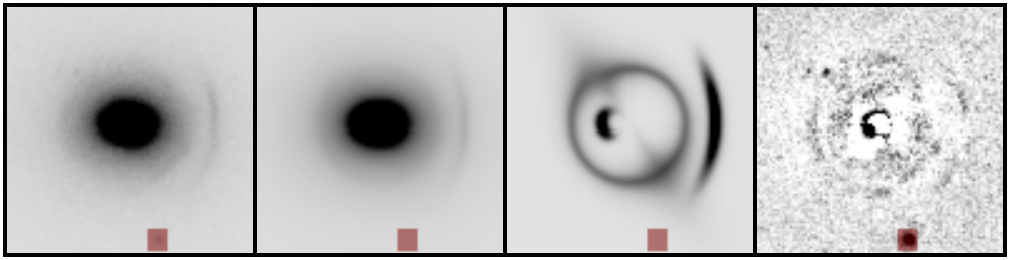}%
}%
\\%
\subfloat[SDSS J0841$+$3824]{%
\includegraphics[width=\columnwidth]{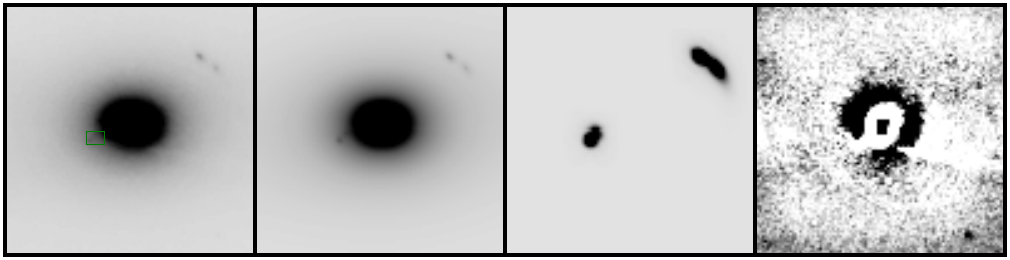}%
}%
\hfill%
\subfloat[SDSS J0903$+$4116]{%
\includegraphics[width=\columnwidth]{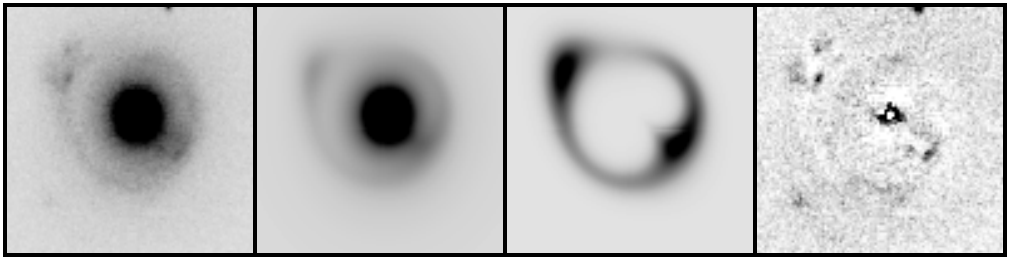}%
}%
\caption{%
Selection of SLACS observations from the F814W band and their reconstructions.
Each reconstruction shows (\emph{left to right}) the original observation, the maximum-likelihood model from \lensed, the lensed images of the background source only, and the residuals.
Images are 5~arcsec by side with north up and east left.
The scale ranges from $-0.25\,X$ (\emph{white}) to $X$ (\emph{black}), where $X$ is the 97th percentile image value of the recovered model for the first and second image, and the 99th percentile image value of the lensed background source for the third and fourth image \citep[cf.\@][]{2008ApJ...682..964B}.
The overlays show masked regions (\emph{red}) or image plane priors (\emph{green}) if present in the reconstruction.
}%
\label{fig:slacs-images}%
\end{figure*}

\begin{figure*}
\includegraphics[width=\textwidth]{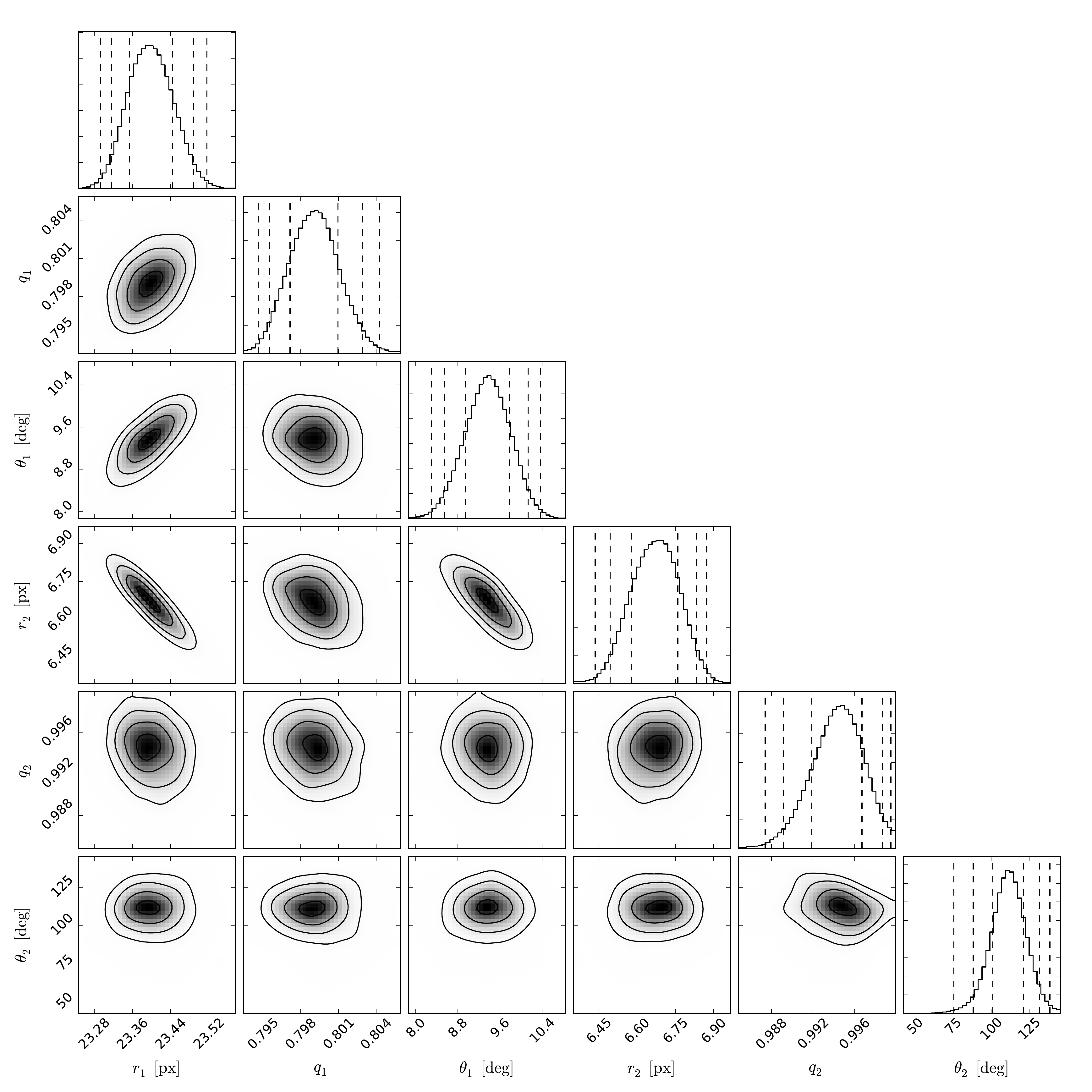}%
\caption{%
Lens parameter constraints for the model of SDSS J0808$+$4706 (Fig.~\ref{fig:J0808+4706}).
The two probable lenses are modelled as SIEs at the same redshift.
The model contains a total of 35 free parameters, including the source and foreground surface brightness distributions.
The triangle plot shows the marginalised one- and two-dimensional posterior distributions for the lens parameters, with contours at the 68\%, 95\%, and 99\% confidence levels.
}%
\label{fig:slacs-posterior}%
\end{figure*}

\begin{table*}%
\centering%
\caption{%
Maximum-likelihood lens models for source reconstruction.
The modelling procedure follows the one used by \citet{2013ApJ...777....1B} as detailed in the text.
The table shows maximum-likelihood parameters for the Einstein radius $b_\text{SIE}$, axis ratio $q$ and position angle of the SIE model, as well as the magnitude $\gamma_\text{ext}$ and position angle of the external shear.
}%
\label{tab:slacs-sources-lenses}%
\begin{tabular}{lcccccccccc}
\hline
& \multicolumn{6}{c}{\hrulefill~~~\lensed~~~\hrulefill}
& \multicolumn{4}{c}{\hrulefill~~~Bandara et al. (2013)~~~\hrulefill} \\
system
& $b_\text{SIE}$~[arcsec] & $q$ & P.A.~(SIE) & $\gamma_\text{ext}$ & P.A.~($\gamma$) & $N_\text{src}$
& $b_\text{SIE}$~[arcsec] & $q$ & $\gamma_\text{ext}$ & $N_\text{src}$ \\
\hline
SDSS J0029$-$0055 & 0.94 & 0.84 & 104.70\degr & 0.028 & 168.90\degr & 2 & 0.94 & 0.86 & 0.03 & 2 \\
SDSS J0252$+$0039 & 1.02 & 0.87 &~~27.71\degr & 0.029 &~~40.23\degr & 3 & 1.03 & 0.90 & 0.02 & 3 \\
SDSS J0330$-$0020 & 1.10 & 0.87 &~~11.02\degr & 0.022 & 137.81\degr & 3 & 1.10 & 0.75 & 0.03 & 3 \\
SDSS J0728$+$3835 & 1.25 & 0.72 & 156.72\degr & 0.055 & 152.91\degr & 3 & 1.23 & 0.74 & 0.06 & 3 \\
SDSS J0822$+$2652 & 1.14 & 0.66 & 155.90\degr & 0.087 & 159.10\degr & 2 & 1.12 & 0.83 & 0.05 & 2 \\
SDSS J0841$+$3824 & 1.45 & 0.66 &~~~~7.73\degr& 0.010 &~~16.63\degr & 2 & 1.40 & 0.78 & 0.01 & 2 \\
SDSS J0903$+$4116 & 1.28 & 0.91 &~~88.17\degr & 0.022 & 137.47\degr & 3 & 1.30 & 0.87 & 0.02 & 3 \\
\hline
\end{tabular}%
\end{table*}

\begin{table*}%
\centering%
\caption{%
Estimated parameters for the S\'ersic source models of 8 SLACS lenses from HST proposal 10886.
The modelling procedure follows the one used by \citet{2013ApJ...777....1B} as detailed in the text.
The table shows the mean and standard deviation obtained from \lensed\ for the effective radius~$r_\text{hl}$, \textit{I}-band AB~magnitude~$m_I$ and S\'ersic index~$n$ of the background components.
Repeated rows marked (a), (b), ... contain results for lens models with multiple source components.
\citet{2013ApJ...777....1B} quote $\sigma_{m_I} \approx 0.38$~mag and $\sigma_n \approx 0.1$ for the total systematic uncertainty of their magnitudes and S\'ersic indices, respectively.
}%
\label{tab:slacs-sources}%
\begin{tabular}{lccccccc}
\hline
&& \multicolumn{3}{c}{\hrulefill~~~\lensed~~~\hrulefill}
 & \multicolumn{3}{c}{\hrulefill~~~Bandara et al. (2013)~~~\hrulefill} \\
system & $z_\text{src}$
& $r_\text{hl}$~[kpc] & $m_I$~[AB] & $n$
& $r_\text{hl}$~[kpc] & $m_I$~[AB] & $n$ \\
\hline
SDSS J0029$-$0055    & 0.93 & 2.688 $\pm$ 0.152 & 23.210 $\pm$ 0.042 & 2.728 $\pm$ 0.100 & 2.72 $\pm$ 0.73 & 23.51 & 2.73 \\
SDSS J0252$+$0039    & 0.98 & 0.814 $\pm$ 0.014 & 23.947 $\pm$ 0.014 & 1.487 $\pm$ 0.039 & 0.98 $\pm$ 0.26 & 24.43 & 1.31 \\
SDSS J0330$-$0020    & 1.07 & 1.678 $\pm$ 0.017 & 21.981 $\pm$ 0.007 & 2.585 $\pm$ 0.035 & 1.83 $\pm$ 0.49 & 22.00 & 2.32 \\
SDSS J0728$+$3835(a) & 0.69 & 2.298 $\pm$ 0.144 & 23.680 $\pm$ 0.057 & 1.809 $\pm$ 0.117 & 2.87 $\pm$ 0.77 & 23.95 & 1.99 \\
SDSS J0728$+$3835(b) & 0.69 & 7.923 $\pm$ 0.565 & 23.180 $\pm$ 0.072 & 3.963 $\pm$ 0.123 & 7.39 $\pm$ 1.98 & 24.31 & 3.83 \\
SDSS J0728$+$3835(c) & 0.69 & 0.104 $\pm$ 0.022 & 27.671 $\pm$ 0.104 & 1.540 $\pm$ 1.055 & 0.11 $\pm$ 0.03 & 27.49 & 0.04 \\
SDSS J0822$+$2652(a) & 1.03 & 1.163 $\pm$ 0.032 & 25.047 $\pm$ 0.410 & 0.361 $\pm$ 0.041 & 1.52 $\pm$ 0.41 & 25.36 & 0.25 \\
SDSS J0822$+$2652(b) & 0.59 & 1.107 $\pm$ 0.032 & 27.789 $\pm$ 0.037 & 0.845 $\pm$ 0.066 & 2.01 $\pm$ 0.54 & 23.96 & 0.72 \\
SDSS J0841$+$3824    & 0.59 & 0.401 $\pm$ 0.013 & 24.454 $\pm$ 0.015 & 0.061 $\pm$ 0.007 & 0.66 $\pm$ 0.18 & 24.43 & 0.16 \\
SDSS J0903$+$4116    & 0.66 & 1.716 $\pm$ 0.015 & 23.627 $\pm$ 0.012 & 0.547 $\pm$ 0.017 & 2.89 $\pm$ 0.78 & 23.51 & 0.55 \\
\hline
\end{tabular}%
\end{table*}

The estimated parameters of our reconstruction are given in Table~\ref{tab:slacs-results}, together with the number of source components used for foreground and background galaxies.
Also listed are the results of the reconstructions by \citet{2008ApJ...682..964B}, which overall are in good agreement with our findings.
The reconstructed images are shown in Fig.~\ref{fig:slacs-images}.
As can be seen from the images, \lensed\ is able to model all individual lensing events, regardless of complexity, signal-to-noise of the background images and overlap between the different sources.
A few cases required particular care due to their peculiar properties; they are briefly described in the following, in order to show the capabilities of the code and possible strategies for modelling non-trivial systems.

\textbf{J0808+4706} --
This system was not modelled in the SLACS paper by \citet{2008ApJ...682..964B} due to the presence of a close-by companion to the main lensing galaxy.
It was qualitatively investigated by \citet{2009MNRAS.399....2O} as an example of a multi-component lens system, but without reconstruction.
With \lensed, we used two separate and independent objects for the main lensing galaxy and its companion, modelling both as SIEs.
We also used two Sérsic components for the light of the main lensing galaxy, one Sérsic component for the companion, and two Sérsic components for the background source.
In total, the model has 35 free parameters, the highest number of all the tests presented in this work.
Despite the complexity of the system, \lensed\ was able to explore the posterior (Fig.~\ref{fig:slacs-posterior}) and obtain reasonable results for the model.
Of particular interest is the strong anti-correlation of the Einstein radii of the two lenses, a feature that would not have been evident in a maximum-likelihood reconstruction method.

\textbf{J0822+2652} --
With the initial model, we were able to get a very accurate reconstruction of the bright arc on the right, but the residuals showed a clear sign of a circular structure around the central galaxy.
As this could be the signature of a second background component, we added it to the model \citep[as did][]{2008ApJ...682..964B}.
In the subsequent reconstruction, the evidence improved significantly ($\Delta \log\mathrm{Ev} \approx 800$).
Quoted are the results from this second model, although we cannot exclude the possibility that the residuals in the first analysis were instead due to a mismatch between the light profile for the central galaxy and the model.
If this is the case, the favoured model is much more elliptical, with an axis ratio of $q_L = 0.6285 \pm 0.0089$.%

\textbf{J0841+3824} --
In this case, most of the lensing information is contained in the very dim counter-image on the bottom left of the host galaxy.
As explained below, we used a mask to make sure that \lensed\ makes use of it.
We note that, even if in the reconstructed image looks similar to the one shown by \citet{2008ApJ...682..964B}, our recovered lens parameters differ significantly from theirs.
This is probably due to the different method used to perform the delicate subtraction of the central galaxy. 


In the case of J0728$+$3835 and J0841$+$3824, the foreground galaxy is not well enough reconstructed by one or two Sérsic components to reduce the significance of its residuals below that of the lensing signal.
Therefore, with a completely free prior distribution for the position, the background source may be placed on top of the central galaxy to make up for the insufficiencies of the foreground model.
This can be prevented by a better modelling of the foreground galaxy.
However, a simple and efficient method is to use \emph{image plane priors} for the background source positions.
Instead of defining the prior distribution on the source plane, one can define a prior on the image plane for the position at which one of the images of the source is required to appear.
Drawing from such a distribution is done in a straightforward way by first sampling a position on the image plane, mapping it to the source plane, and assigning it to the source.
This is a ``prior'' in the true sense of the definition, as it uses an observer's prior knowledge regarding the apparent position of an image for the distribution of possible source positions.
The use of image plane priors reduces the correlations between lens and source parameters, as a source normally has to be repositioned any time the lens is updated to make the observed images appear once again where they are observed.
At the same time, the parameter space volume of the source position is dramatically reduced, as the observed position is well constrained from the start.
The image plane prior regions we chose in the two cases are visible in Fig.~\ref{fig:slacs-images}.

As evidenced by the reconstructions of this section, \lensed\ is able to obtain robust and meaningful results from the analysis of real data, where the uncertainties are much more severe than those of the mock data analysed in the previous section.
The algorithm is able to reconstruct at the same time the light coming from the central galaxy and from the background lensed sources.
As in the analysis of mock data, no informative priors were used, and the analysis was, apart from the model choices described above, fully automatic.
The few cases in which there are relevant discrepancies with previous analyses, these are due to inherent difficulties in the identification of lensing signatures, such as the circular residual in J0822$+$2652 and the poor counter-image in J0841$+$3824.
These are peculiar cases that inevitably require some decision made by the modeller on the interpretation of the lensing event. In all cases in which there are clear signs of multiple images of the background source, \lensed\ is able to appropriately disentangle them from the light of the host galaxy, even in cases in which there is significant overlap (J0330$-$0020, J0903$+$4116).
It should be clarified that the formal errors quoted in this analysis are purely statistical, and correspond to real uncertainties only in the ideal case in which the model used for lenses and sources offers a correct description of these objects. In the analysis of real data, this may not be completely true, and model-free quantities such as the Einstein radius of the lens can be affected by the uncertainty on the model. Thus, if one wants to interpret the quoted values in an extensive way, an additional empirical error of the order of a few percent should be considered, as done in other works \citep{2008ApJ...682..964B,2013ApJ...777...97S}.

Finally, we wish to briefly comment on the reconstruction of background sources with \lensed.
There is only limited room for comparison of our sample with the results of \citet{2011ApJ...734..104N}, as only one of our systems has been analysed there.
Instead, we look at the results of \citet{2013ApJ...777....1B}, where we find all our systems except J0808$+$4706 (for the same reason as above).
To obtain comparable results, we perform a new analysis of our systems following a similar reconstruction procedure.
In a first step, we model the lens systems using a suitable number of foreground components, a SIE lens with external shear, and the same number of background components as \citet{2013ApJ...777....1B}.
The maximum-likelihood parameters we obtain are shown in Table~\ref{tab:slacs-sources-lenses}.
We find that the results agree both with our previous analysis and the SLACS results quoted above.
The differences in alignment and ellipticity are explained by the presence of external shear \citep[see the discussion in][]{2013ApJ...777....1B}.
More importantly, our results are in good agreement with those obtained by \citet{2013ApJ...777....1B}, except for the position angles of the mass model and external shear.
In light of the consistency of our results, we conclude that there is a problem with their reported position angles,\footnote{%
If one (incorrectly) assumes that the individual images of our systems in the extended Fig.~3 of the online version of \citet{2013ApJ...777....1B} are north-aligned, the position angles appear to be in agreement.}
which are therefore not reproduced here.
Of note is the system J0822$+$2652, for which we find a more elliptical lens ($q = 0.66$) and a strong external shear ($\gamma_\text{ext} = 0.09$).
Considering that the signal of the second source was only identified as a ring-like structure in the residual image (see above), it is likely that there is a pronounced degeneracy between ellipticity and external shear, and even subtle differences in the reconstruction can lead to vastly different results in the degenerate $q$~--~$\gamma$ plane.
This is corroborated by the close alignment of the SIE and external shear position angles.
Besides, the level of agreement between the codes is generally high, particularly in light of the fact that our lens models are not built iteratively component after component as in \citet{2013ApJ...777....1B}, but instead in one pass.

Once the lens systems have been modelled, \citet{2013ApJ...777....1B} fix the foreground and lens parameters to their maximum-likelihood values, and the individual background components are replaced by a single source.
They perform this second step in order to obtain a ``global'' S\'ersic profile model of the source plane.
(This second step is not performed for systems with backgrounds classified as having ``group'' morphology.)
Using \lensed, we are able to obtain these results by simply removing all but one background source components, except in the case of J0841$+$3824.
This system shows a strongly bimodal posterior distribution, in which the preferred solution reconstructs only one of the two sources.
After suitably reducing the allowed parameter space, we obtain the desired ``global'' solution which envelopes both background source components.
The full set of our results are shown in Table~\ref{tab:slacs-sources}, where we once again find good agreement between the lens reconstruction codes.
Differences for the radii, magnitudes and S\'ersic indices of the reconstructed sources are reasonable, given the indicated errors.
In the case of J0822$+$2652, we obtain smaller and dimmer sources, which can be explained by differences in magnification for our more elliptical, strongly sheared lens model.
Nevertheless we find acceptable agreement for the S\'ersic indices, and hence the shape of the source profile, in this system, as in the others.

Based on this preliminary analysis, we conclude that \lensed\ is able to recover the properties of background sources.
However, the size of our sample is too small to draw conclusions about whether there are intrinsic differences in the reconstruction of lensed sources when different algorithms are used.
In the future, it will become necessary to perform a thorough and systematic comparison of lensing codes for the analysis of lensed sources.

\section{Conclusion}

We have introduced \lensed, a new code for the forward lens reconstruction from a parametric model.
The forward method, presented in section~\ref{sec:fwdrec}, offers conceptual simplicity and allows for a clear interpretation of the results for both lenses and sources.
It is furthermore an important tool for testing new analytical models of lenses and -- eventually -- sources, at higher redshifts than otherwise accessible.

There are however significant challenges that must be overcome in any truthful implementation.
First, if the simulation of the assumed model is not accurate, i.e.\@ an insufficient integration of the surface brightness distribution for each observed pixel, there will inevitably be a number of systematic biases in the results, as the quality of the simulation degrades close to the areas of high magnification, and thus high signal.
Secondly, the parameter space for a lens reconstruction problem is usually highly degenerate with multiple local maxima of possible solutions, making a straightforward sampling very difficult.

In section~\ref{sec:implementation}, we have listed our solutions to the most severe of these problems.
By carefully rewriting calculations in a massively parallel fashion, we were able to harness the multiprocessing capabilities of modern GPUs and maximise the efficiency of our code.
The resulting increase in computational speed allowed us to simulate the expected light distribution using numerical quadrature for each individual pixel, and still test thousands of individual parameter settings per second.
Coupling the fast and accurate simulation with a modern library for Bayesian analysis made it possible to achieve a true sampling of the system's parameter space in reasonable time.

Section~\ref{sec:testing} contains the tests we have performed to ensure our implementation is indeed working as expected.
This has been done by analysing a fixed lens model in a sample of 100 mock images containing randomised sources.
Recovering the lens parameters in a setting where the model can be exactly replicated in \lensed, we were able to show that no biases are introduced in the reconstruction.
This remains true after switching to more complex background galaxies based on observations, and thus an inexact model within \lensed.
Even in this case, we recovered the true values to high accuracy.
Thus we are confident that in real applications, the quality of the results is not limited by the algorithm we are using for the reconstruction.

Finally, we used \lensed\ to analyse a number of real observations in section~\ref{sec:slacs}.
The data originates from the SLACS survey of \citet{2006ApJ...638..703B,2008ApJ...682..964B} and has been previously analysed there.
We took this opportunity to compare our reconstructions to published results in the literature, highlighting eventual differences and their origins, while also showcasing how \lensed\ can be used in practise.

There are a number of possible extensions to the algorithm in its current form.
One such extension is support for the multi-plane lensing formalism of \citet*{2014MNRAS.445.1954P}, and \lensed\ was designed with this goal in mind.
Another possibility is the simultaneous reconstruction of observations in multiple bands \citep{2013MNRAS.435..623V,2014MNRAS.440.2013D,2014MNRAS.441.3238K}.
Due to the speed and robustness of the implementation, it is also conceivable to use \lensed\ in a lens finding scenario such as the one recently proposed by \citet{2015A&A...577A..85B}.
Finally, it is possible to implement specialised treatment for lensing by galaxy clusters, the images of which are much larger in size than those of galaxy-galaxy lensing events, offering unique opportunities for optimisation.

\section*{Acknowledgements}

This research is part of project GLENCO, funded under the European Seventh Framework Programme, Ideas, Grant Agreement n.~259349.

The analysis of section~\ref{sec:slacs} is based on observations made with the NASA/ESA Hubble Space Telescope, and obtained from the Hubble Legacy Archive, which is a collaboration between the Space Telescope Science Institute (STScI/NASA), the Space Telescope European Coordinating Facility (ST-ECF/ESA) and the Canadian Astronomy Data Centre (CADC/NRC/CSA).

We would like to thank D. Leier for his testing and feedback during the development of the code, and the referee for their thoughts on the reconstruction of lensed background sources.



\bibliographystyle{mnras}
\bibliography{lensed}


\bsp	
\label{lastpage}
\end{document}